%% file: 0_main.tex
\documentclass{vgtc}                          

\newcommand{\dhk}[1] 
{\textcolor{black}{#1}}
\newcommand{\dhkk}[1]
{\textcolor{black}{#1}}
\newcommand{\ic}[1] 
{\textcolor{black}{#1}}
\newcommand{\hdy}[1] 
{\textcolor{black}{#1}}
\newcommand{\noSig}[1] 
{\textcolor{lightgray}{#1}}

\graphicspath{{figures/}{pictures/}{images/}{./}} 

\usepackage{times}                     

\usepackage{tabu}                      
\usepackage{booktabs}                  
\usepackage{lipsum}                    
\usepackage{mwe}      

\usepackage{mathptmx}                  
\usepackage{multirow}

\onlineid{1914}

\vgtccategory{Research}

\vgtcinsertpkg

\title{Crossing Rays: Evaluation of Bimanual Mid-air Selection Techniques in an Immersive Environment}

\author{
DongHoon Kim
\thanks{e-mail: donghoon.kim@usu.edu}\\
\scriptsize Utah State University
\and Dongyun Han
\thanks{e-mail: dongyun.han@usu.edu}\\ %
\scriptsize Utah State University %
\and Siyeon Bak
\thanks{e-mail: M23051@hallym.ac.kr}\\
\scriptsize Hallym University
\and Isaac Cho
\thanks{e-mail: isaac.cho@usu.edu}\\ %
     \parbox{1.4in}{\scriptsize \centering Utah State University}}

\teaser{
  \centering
  \includegraphics[width=\linewidth]{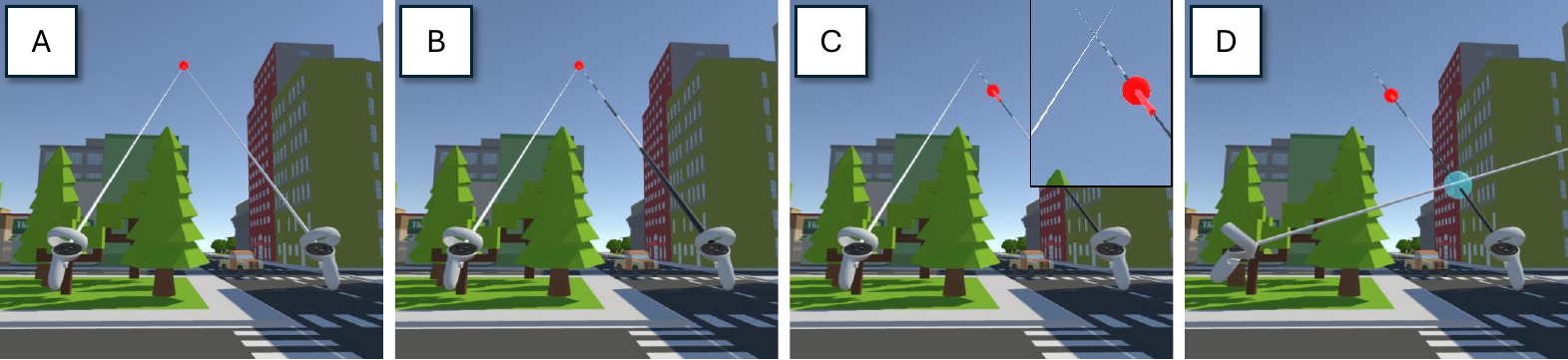}
  \caption{
  A: Simple-Ray: Mid-air selection by intersecting two straight rays projected from the controllers held in each hand.
  B: Simple-Stripe: Alternating black and white stripes, which measures 1m, along the ray in the dominant hand.
  C: Precision-Stripe: An extra step where a stripe is selected to minimize the likelihood of unintended cursor movements.
  D: Cursor-Sync: Maneuvering the cursor within the first stripe of the dominant ray for precise control.
  }
  \label{fig:teaser}
}

\abstract{
Mid-air navigation offers a method of aerial travel that mitigates the constraints associated with continuous navigation.
A mid-air selection technique is essential to enable such navigation.
In this paper, we consider four variations of intersection-based bimanual mid-air selection techniques with visual aids and supporting features: Simple-Ray, Simple-Stripe, Precision-Stripe, and Cursor-Sync.
We evaluate their performance and user experience compared to an unimanual mid-air selection technique using two tasks that require selecting a mid-air position with or without a reference object.
Our findings indicate that the bimanual techniques generally demonstrate faster selection times compared to the unimanual technique. With a supporting feature, the bimanual techniques can provide a more accurate selection than the unimanual technique.
Based on our results, we discuss the effect of selection technique's visual aids and supporting features on performance and user experience for mid-air selection.
} 

\keywords{Virtual Reality, Mid-Air Selection, Bi-manual Interaction}

\begin{document}



\maketitle

\input{ANOVA_Result_contents/0_introduction}
\input{ANOVA_Result_contents/1_related_works}
\input{ANOVA_Result_contents/2_techniques}
\input{ANOVA_Result_contents/3_experiment}
\input{ANOVA_Result_contents/4_result_more_short}
\input{ANOVA_Result_contents/5_discussion}

\input{ANOVA_Result_contents/6_conclusion}

\bibliographystyle{abbrv-doi}

\bibliography{reference}
\end{document}

%% file: ANOVA_Result_contents/0_introduction.tex
\section{Introduction}

Selection represents a core interaction within immersive virtual environments (IVE), playing a crucial role in both object manipulation and the facilitation of several navigation techniques~\cite{argelaguet2013survey}.
For example, Teleportation~\cite{prithul2021teleportation} allows users to move to a desired position immediately, but the user must select a target position.
However, these techniques require the selection of a tangible object in IVE, such as a ground or a virtual object to move close to, which limits users' freedom of navigation. 
Importantly, the scope of navigation is not confined to ground level, allowing a comprehensive spatial interaction that encompasses various elevations and dimensions \cite{weissker2023gaining}.
Thus, a mid-air selection technique, which allows selection without any tangible object, can make navigating into the sky possible.
Mid-air selection can be defined as a technique that enables the user to point and select an intangible object within a 3D space.
Previous studies have utilized scene-in-hand metaphor \cite{ware1990exploration} along with arm-extension techniques \cite{li2018evaluation, cho2015evaluation}, a parabolic ray starting from a ground position, or combining keys to control different degrees of freedom (DOF) \cite{weissker2023gaining} to facilitate mid-air selection.
Nonetheless, these approaches have restrictions such as constraints being on the ground, a limited range of selections, and inaccurate position selection due to the sensitivity of offset gain factors \cite{li2018evaluation}.

\hdy{We identify and consider the four variants of mid-air selection techniques.}
\hdy{A basic bimanual technique, Simple-Ray, uses two linear rays emanating from both hands to create an intersection point and use the point as a selecting position. }
On top of this Simple-Ray technique, a visual aid and selection mechanisms are added to enhance the selection accuracy and user experience.
Simple-Stripe has 1-meter length stripes on a ray. Precision-Stripe allows the user to select one of the 1-meter length stripes and making the intersection point only be located within the selected 1 meter.
Cursor-Sync allows the user to create an intersection point inside the stripe located closest to the controller.  

We evaluate the performance of these techniques in two types of mid-air selection tasks with 20 participants.
The first mid-air selection task is to select a mid-air position near an object distantly located in front of a participant to evaluate the performance of selecting a given mid-air position.
The second selection task is to select a mid-air position where a target object was located for 1.5 seconds.
This task evaluates the performance of selecting a mid-air position that a user wants to select.
Both tasks are performed at three different distances (3m, 6m, and 9m) to investigate how performance varies by distance.
The results show that Simple-Ray allows a user to select a position quickly.
Unexpectedly, Simple-Stripe represents a not different selecting time and accurate results both from Simple-Ray results.
Precision-Stripe has relatively better accuracy at 6m and 9m distances, but the selection time is the slowest among the evaluated bimanual techniques.
Interestingly, Cursor-Sync allows participants to select a mid-air position significantly faster than Precision-Stripe. However, different from the expectation, the accuracy is not significantly different.
The One-Hand mid-air selection technique~\cite{weissker2023gaining} that is evaluated for comparison generally shows inaccurate and slow selections compared to bimanual techniques.
\hdy{The main contribution of this work is identifying the mid-air selection techniques and evaluating them for users which can ultimately be used for a better mid-air navigation experience in various VR environments through these techniques. }



%% file: ANOVA_Result_contents/1_related_works.tex
\section{Related Works}
Previous studies introduced selection techniques that can be used for mid-air selection. In this section, we introduce them by explaining how they are designed and how they work~\cite{bowman1999testbed, hinckley1994survey}.

\subsection{Mid-Air Selection}
\paragraph{Direct Selection}
Direct selection with a controller is the primary method for selecting objects in VR, similar to grabbing objects in your hand in the real world~\cite{jacoby1994gestural,mine1997moving}.
It allows users to perform mid-air selection within their arm's reach space.
However, it has a limitation of selecting distant coordinates beyond their arm reach~\cite{song1993nonlinear}.
To address such a limitation, selection techniques for distant objects have been developed, some of which can be adaptable to mid-air selection.
An example method is the arm extension metaphor, including Go-Go and the linear offset technique~\cite{poupyrev1996go, bowman1997evaluation}.

These techniques entail the mapping of virtual cursors or hands exponentially or with a designated offset, contingent on the distance between the user and the controllers.
It enables virtual cursors to extend beyond the actual controller position.
However, if the arm extension ratio is increased to select far away, it decreases selection performance, such as selection time and accuracy~\cite{li2015evaluation}.
Another approach to solving arm reach limitation is decreasing the size of the world~\cite{pausch1995navigation,stoakley1995virtual, bell2001view}.
With this method, a user can directly select a position, even a mid-air position, from a different viewpoint, as if looking down from the sky.
Also, this method not only extends the selectable area but also presents the occluded position behind objects.
However, the different scale viewpoints can make selecting a position hard for a user, and when the scale ratio is changed, the difficulty can be increased~\cite{wingrave2006overcoming}.

\paragraph{Ray Selection}
Next, the ray casting metaphor, in which rays emerge from controllers, eyes, or the head, is a popular method for remote object selection.
Diverse ray casting techniques have been introduced to achieve accurate and fast distant object selection in a variety of scenarios.
For example, Lu et al.~\cite{lu2020investigating} introduced 3D Bubble Cursor, which uses a heuristic approach to provide a sphere cursor that dynamically adjusts its size to always capture the closest target from the ray.
Steinicke et al.~\cite{steinicke2006object} presented the Sticky-Ray selection technique, which is based on a bendable ray that points to the closest selectable object.
However, many of these variations still require target objects and do not support mid-air selection as they only take into account 2 DOF of pitch and yaw controller rotation for remote object selection.
A way to make the ray casting technique enable mid-air selection is additional degrees of freedom to control the depth of the ray by a ray that has been manipulated using multimodal input.
Previous research incorporated depth control into the ray casting metaphor to address remote object selection issue in occlusion. 
Baloup et al. introduced RayCursor~\cite{baloup2019raycursor}, allowing depth adjustment of ray selection by adding a cursor that can move forward and backward using controller buttons.
Ro et al.~\cite{ro2017dynamic} proposed a similar approach using a smartphone.
Similarly, Zhai et al.~\cite{zhai1996partial} presented Silk Cursor, a 3D volume cursor rather than a 3D point cursor.
A ray can also select a mid-air position by changing the height of a cutoff layer.
Parabola teleportation is the most common teleportation method that selects a position on the ground using a curved ray~\cite{funk2019assessing}.
In that case, the curved ray is cut at the ground level to select a position on the ground, but when the height of the cutoff layer is changed, this curved ray can now select the mid-air position~\cite{weissker2023gaining}.
\dhkk{
This elevating the ground-level method was also studied by lifting an object from a floor and positioning it in mid-air~\cite{sun2019extended}.
}

\subsection{Bimanual Selection}
Another multimodal approach is using two hands.
Using a bimanual technique on a single task can provide better performance than the usage of an unimanual technique~\cite{buxton1986study}.
Owen et al.~\cite{owen2005gets} suggested design principles for a better bimanual technique than unimanual: visually integrated, conceptually integrated, and integrated device spaces.
And this improvement is more effective when a task gets more difficult.

Several studies tried to improve the unimanual technique using the bimanual approach.
Ulinski et al.~\cite{ulinski2007two,ulinski2009selection} had a volumetric data interaction study focused on the technique's symmetricity.
Their study suggested a design principle for the bimanual selection technique: a symmetric technique can increase selection speed, and an asymmetric structure can decrease fatigue.
Chaconas et al.~\cite{chaconas2018evaluation} investigated a bimanual manipulation technique in an augmented reality environment.
Their technique can have rotation and scale operations using a bimanual gesture system without controllers, and its user study results indicate that the bimanual technique can have a place in the future of augmented reality.

Also, the bimanual technique can increase the number of DOF~\cite{cho2015evaluation,feng2015comparison,li2015evaluation}.
While a user controls the two 6 DOF controllers, the relative position between the controllers allows them to handle the additional DOF, resulting in a consequent 7 DOF system.
The additional DOF was used to control the scale of a selected object intuitively in the studies, but it still has the potential to be used as various features, such as tension control.
For instance, a study focusing on indicating a mid-air position used a bimanual control system to control an arrow's length and curvature that indicated a mid-air position~\cite{feiner2003flexible}.

The bimanual approach also applies to the ray casting based technique.
This approach addresses the limitations of unimanual ray casting by distinguishing between quick but rough and fine selection tasks.
The fine selection activity gives users control over the depth of ray selection, resulting in mid-air selection.
Wyss et al.~\cite{wyss2006isith} and Zhang et al.~\cite{zhang2022conductor} introduced bimanual ray casting techniques called iSith and Conductor, respectively.
iSith is an intersection-based selection technique that employs two rays emerging from both hands, whereas Conductor uses one ray for the dominant hand and a rectangle-like wide ray for the minor hand.
Zhang et al. conducted a user study and found that the Conductor outperformed the RayCursor in remote target selection, but did not compare its performance with that of iSith.
Chen et al.~\cite{chen2023gazeraycursor} introduced GazeRayCursor, a multimodal ray control approach that uses both eyes and controllers.
Although these intersection-based bimanual techniques have the potential for selecting mid-air positions, they were primarily studied for object selection which requires a selectable object.
Therefore, it is challenging to use these techniques to select a mid-air position without selectable objects. Consequently, we excluded these techniques from our user study. 

In this paper, we investigate the intersection-based bimanual technique for mid-air selection.
Additionally, using the distance conditions (near, medium, and far away), we investigate the potential use of it as mid-air navigation's destination selection technique.

%% file: ANOVA_Result_contents/2_techniques.tex
\section{Crossing Rays for Mid-Air Selection}
A recognized challenge with ray-based selection techniques is the difficulty encountered when the target is situated at a distance, which is known as the Heisenberg effect \cite{wolf2020understanding}.
This issue persists in the context of mid-air selections, affecting the precision and ease of target acquisition from afar.
\hdy{Considering this challenge, this work evaluates the following four bimanual} ray-based selection variants -- namely \textbf{Simple-Ray}, \textbf{Simple-Stripe}, \textbf{Precision-Stripe}, and \textbf{Cursor-Sync}. \hdy{We call them Crossing Rays in this work.}
Furthermore, we detail the \textbf{One-Hand Simultaneous} selection technique introduced by Weissker et al.~\cite{weissker2023gaining} as a technique utilized in our user study to assess the efficiency and usability of \hdy{Crossing Rays}.

\begin{figure*}[t]
\centering
\includegraphics[angle=0, width=.95\textwidth]{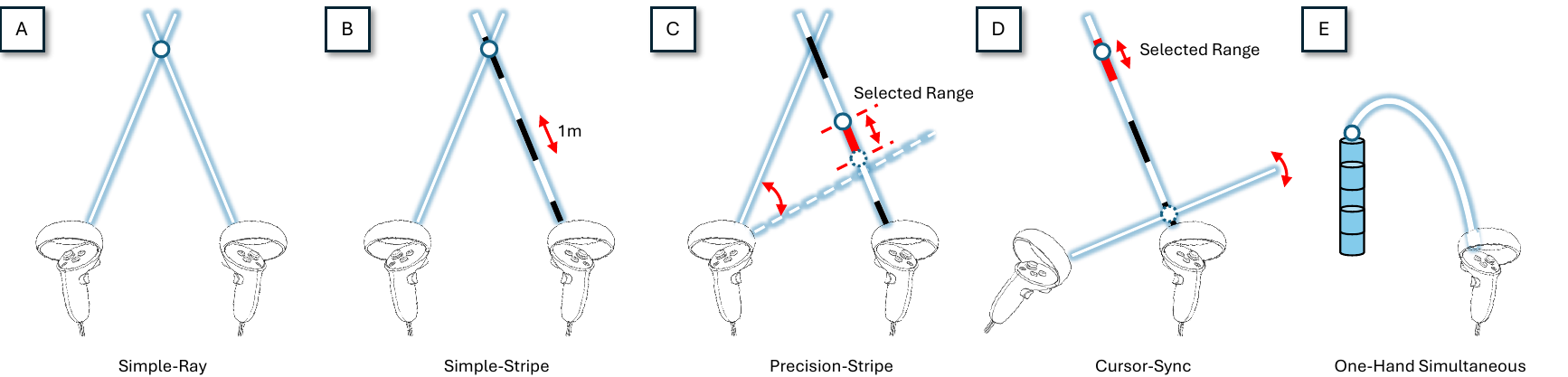}
\caption{
Crossing Rays:
A) \textbf{Simple-Ray}: Two rays from each controller make an intersection point, pointing to a mid-air position;
B) \textbf{Simple-Stripe}: The base is the Simple-Ray technique, and 1-meter length stripes are on the ray of the dominant hand side;
C) \textbf{Precision-Stripe}: The intersection point can only be located in the selected range, one of the stripes; and 
D) \textbf{Cursor-Sync}: The selecting point's relative position in the selected range is synchronized with the position of the intersection point in the closest stripe.
E) \textbf{One-Hand Simultaneous}: The unimanual mid-air selection with parabolic ray, which can control vertical position using a joystick on the controller pushing (upward) and pulling (downward).
} 
\label{figure:techniques}   
\vspace{-0.2cm}
\end{figure*}

\subsection{Basic Crossing Ray Technique: Simple-Ray (SR)}
The Crossing Ray technique allows mid-air selection by intersecting two straight rays projected from the controllers held in each hand, a mechanism introduced by Wyss et al. \cite{wyss2006isith}.
By default, it shows two simple white rays.
Users can control the intersection point by altering the orientation of the controllers, directing them inward or outward.
Additionally, they can adjust the intersection point by moving the controllers either closer to or farther.
In accordance with Guiard's framework \cite{guiard1987asymmetric}, the ray emitted from the dominant hand targets the desired point, while the ray from the non-dominant hand establishes a reference frame for the dominant hand, facilitating coordinated action and spatial orientation.

When two rays are interested, a sphere cursor is visible, as shown in Figure~\ref{figure:techniques}A.
The cursor is a 0.1m radius white sphere.
The cursor appears when the two rays are less than 0.2m apart orthogonally, even if they do not intersect perfectly.
This design choice aims to minimize user's arm fatigue and simplify the technique's utilization \dhk{rather than intersecting exactly}.

\subsection{Simple-Stripe}
\dhk{To evaluate the effect of visual aid, this variant} presents a visual aid of alternating black and white stripes along the ray projected from the controller in the dominant hand, as shown in Figure~\ref{figure:techniques}B.
Each black or white stripe measures 1m.
By integrating stripes to function as a visual ruler, this \dhk{variant technique} is expected to enhance the user's ability to perceive depth and accurately determine mid-air position within the virtual space.
The incorporation of these stripes into the ray selection process introduces a novel layer of user interaction by providing a clear, measurable reference for assessing spatial information in the virtual environment.

\subsection{Precision-Stripe}
This \dhk{variant} builds upon the Simple-Stripe by incorporating an extra step where a stripe is selected prior to the final mid-air selection.
The selection of a stripe is facilitated through the joystick of the dominant hand's controller.
\dhk{
Pushing the joystick forward moves the selection further, and pulling it backward moves it closer.
}
This modification enables users to confine the scope of mid-air selection, thereby minimizing the likelihood of unintended cursor movements—either further away or closer—due to fluctuations in the controller's position (Figure~\ref{figure:techniques}C).
This feature could be particularly beneficial when the target is positioned at a distance from the users.

\subsection{Cursor-Sync}
Cursor-Sync \dhk{variant} empowers users to maneuver the cursor within the first stripe of the dominant ray proximal to them, as illustrated in Figure~\ref{figure:techniques}D.
Initially, upon selecting a stripe of interest, two cursors materialize.
One within the chosen stripe and another in the first stripe of the dominant ray.
The positions of these cursors within their respective stripes are synchronized.
Subsequently, when users adjust the position of the near-cursor using the non-dominant ray, the cursor within the selected range of interest correspondingly shifts.
With this technique, users could precisely control the selection point, enhancing accuracy and interaction efficiency.

\subsection{One-Hand Simultaneous}
Weissker et al. introduced the One-Hand Simultaneous technique for mid-air teleportation \cite{weissker2023gaining}.
Users can utilize the One-Hand Simultaneous (One-Hand) technique through a touchpad to adjust their target position vertically while concurrently moving the controller to pinpoint the desired location.
Their research highlights the advantages of the One-Hand teleportation technique compared to the other two methods.
The Two-Step technique enables users to choose a target location at the same elevation and adjust the controller's tilt to move vertically.
Conversely, the Separate technique requires users to modify the elevation first, followed by teleportation within that same elevation level.
The findings indicate that the simultaneous technique achieves the highest accuracy, despite resulting in longer specification times and increased task load compared to the other two techniques.

In our study, we implemented that simultaneous technique as the One-Hand technique utilizing a parabolic ray with a cursor at its end (Figure~\ref{figure:techniques}E).
\dhk{Moreover, due to this technique has} straightforward control mechanism and higher accuracy for mid-air selection, we employed this as a baseline condition \dhk{for comparing to the Crossing Ray techniques.} 

%% file: ANOVA_Result_contents/3_experiment.tex
\begin{figure}[t]
\centering
\includegraphics[angle=0, width=.47\textwidth]{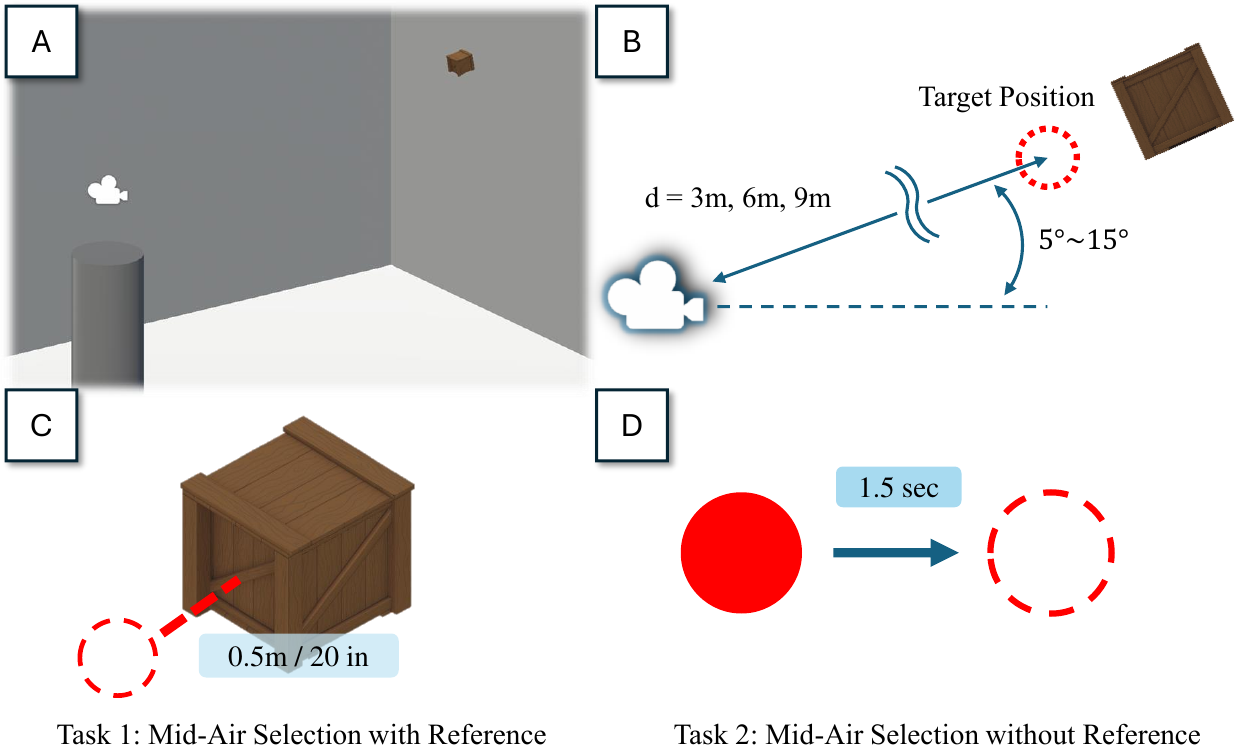}
\caption{
A) \textbf{Experiment room} is a 15m width, depth, and height empty space with a 5m height pillar for a participant.
B) \textbf{Target position} is located in front of the participant with three \dhk{different} distance conditions (3m, 6m, and 9m).
The location is 10 degrees away from the central axis.
C) In \textbf{Task 1}, the target position is 50cm in front of a reference object (Wooden box).
The participant can only select within 50cm of the target position.
D) In \textbf{Task 2}, a red sphere is a target position. \hdy{It disappears 1.5 seconds after the start of each task.}
\dhk{
After the target has disappeared, the participant has to select the position where the red sphere was using a given technique.
}
} 
\label{figure:experiments}   
\vspace{-0.2cm}
\end{figure}

\section{Experiment design}
In this section, we outline the design of our user study aimed at assessing the usability and efficiency of the mid-air selection techniques described in Section 3.

\subsection{Experiment Tasks}
For the study, we crafted two mid-air selection tasks to accommodate two distinct scenarios within a VR environment.
The initial scenario involves selecting a mid-air position adjacent to a virtual object, whereas the subsequent scenario entails choosing a mid-air position in the absence of reference objects.

\subsubsection{Task 1: Mid-Air Selection with Reference} 
This task involves selecting a position located 0.5 meters in front of a reference object suspended in mid-air.
The reference object, a wooden box, is oriented such that its facing side is perpendicular to the user (Figure~\ref{figure:experiments}C). 
The target position varies across three different distance conditions (3m, 6m, and 9m) and is positioned 10 degrees away from the central axis.
Participants are required to use a specific selection technique to identify the target position.
Selections are only valid if made from a minimum distance of 50 centimeters away from the target location.
Participants click a button and the position does not align with this threshold, they must click again until the selection criteria are met.

In this task, the presence of a reference object eliminates the need for participants to remember the mid-air position, allowing them to concentrate solely on selecting the designated location.
The measurement focuses on the selection time required for a participant to accurately select the correct position over multiple attempts.
Therefore, the selection time reflects the participants' ability to quickly identify a strictly designated object or position.
Meanwhile, the selection accuracy serves as an indicator of how effectively a technique assists users in gauging distance from a nearby object.

\subsubsection{Task 2: Mid-Air Selection without Reference}
This task requires selecting a mid-air position previously occupied by a red sphere.
The red-sphere is at one of three distinct distance conditions (3m, 6m, and 9m) and offset by 10 degrees from the center.
The sphere is displayed for 1.5 seconds at the start of the task, after which participants are required to recall and select the sphere's position once it has disappeared.

During the red-sphere is visible, only one ray from the dominant hand is visible for the crossing rays techniques \dhk{for evaluating mid-air position perception using visual aid. }
In contrast to Task 1, the absence of a reference object necessitates that users remember the designated position, challenging them to select based on memory, which can be viewed as a measure of intentional position selection.

This setup evaluates each technique's effectiveness in facilitating a user's ability to intentionally select a mid-air position.
Selection accuracy measures how well a technique assists in perceiving and remembering a target position.

\subsection{Measurement}
For both Task 1 and Task 2, \dhk{two} quantitative metrics are recorded to assess user performance: \textbf{selection time} and \textbf{error distance}. 
Selection time measures how quickly the participant performs the selection task with or without a reference object.
Error distance calculates the discrepancy between the target position and the participant's selected position.

Following each technique and task, participants complete two questionnaires: the \textbf{NASA-TLX}, which assesses six categories (Mental Workload, Physical Workload, Temporal Load Operation, Performance Capability, Effort, and Sense of Frustration) as detailed by \cite{hart1988development,hart2006nasa}, and the \textbf{User Experience Questionnaire} (UEQ) for evaluating six different aspects of user experience including (Attractiveness, Perspicuity, Efficiency, Dependability, Stimulation, and Novelty) as outlined by \cite{laugwitz2008construction}.
These instruments gauge the participants' VR experiences with the respective technique and task.

\subsection{Procedure}
When a participant arrives at a study location, he or she is asked to read and sign a consent form according to the IRB protocol (\#13746).
The participant then completes a demographic questionnaire about the participant's age and VR experience.
Following that, the participant is briefed on the experiment's purpose, procedure, and tasks.
The participant practiced each technique and selection task right before the experiment with that technique and task type condition.
Each participant performs a total of 180 selection tasks (2 tasks \texttimes{} 5 techniques \texttimes{} 3 distances \texttimes{} 6 times repeat).
The technique order is counterbalanced across all participants, while the distance is randomly ordered for the tasks.
After completing a task using a given technique under a task type, the participant is asked to complete the NASA-TLX and UEQ questionnaires.

\subsection{Hypothesis}
For this study, we set the following five hypotheses.

\begin{itemize} 
\item[\textbf{H1}] Simple-Ray and Simple-Stripe are expected to have faster selection time than the others as they do not incorporate an additional clutch that could potentially delay selection.

\item[\textbf{H2}] Cursor-Sync is expected to have a faster selection time than Precision-Stripe and One-Hand because it allows for more precise selection and reduces time spent on detail control.

\item[\textbf{H3}] Precision-Stripe and Cursor-Sync are expected to have higher accuracy compared to the other techniques due to visual aids and support features that facilitate precise selection. 

\item[\textbf{H4}]  Simple-Stripe is expected to have higher accuracy than Simple-Ray and One-Hand due to visual aid.

\item[\textbf{H5}] Simple-Stripe, Precision-Stripe, and Cursor-Sync are expected to provide better usability and less workload than Simple-Ray and One-Hand because of the visual aid and support feature.
\end{itemize}

\subsection{Participants}
A total of 20 participants (7 males and 13 females) are recruited from the university's participant recruitment system (SONA).
The minimum participant number to ensure the statistical effect size is 16. 
The number is calculated using the G*power analysis tool~\cite{faul2007g} with the following parameters: effect size = 0.4, alpha = 0.05, and power = 0.9.
Their average age is 19.3, with a range of 18 to 22 years. All participants have 20/20 (or corrected 20/20) vision, and they do not have physical impairments in using VR devices.
They were rewarded 2.0 SONA credits as follows the university SONA policy.
According to the pre-questionnaire, 13 out of 20 participants have VR experience. 

\subsection{Apparatus}
The Meta Quest 3 and its two controllers are utilized.
It features a 110$^\circ$ horizontal and 96$^\circ$ vertical Field of View (FoV), and its resolution is 2064 \texttimes{} 2208 pixels per eye.
We present the target position located within 60$^\circ$ FoV in all conditions.
The experiment application is developed by Unity 2022.3.3f1 and runs on a Windows 10 desktop with an Intel Xeon W-2245 CPU (3.90GHz), 64GB RAM, and Nvidia GeForce RTX 3090 graphics card.

%% file: ANOVA_Result_contents/4_result_more_short.tex
\section{Results}

\begin{table}
\centering
\caption{
Task 1 - ANOVA Analysis Results.
The shaded rows represent results that have no significant difference.
}

\begin{tabular} {m{1.43cm}| m{0.7cm} | m{0.3cm} | m{0.54cm} | m{0.457cm}}
\toprule 
\multicolumn{5}{c}{\textbf{Selection Time}}\\ \midrule

    {\textbf{}}&\multicolumn{1}{c|}{df} & \multicolumn{1}{c|}{F} & \multicolumn{1}{c|}{\textbf{$p$}} & \multicolumn{1}{c}{\textbf{$\eta$\textsubscript{$p$}\textsuperscript{$2$}}} \\ \midrule

    \multicolumn{1}{c|}{\textbf{Tech. $\times$ Dist.}} &\multicolumn{1}{c|}{8, 152}& \multicolumn{1}{c|}{7.247}    &   \multicolumn{1}{c|}{\textless .001}    &   \multicolumn{1}{c}{.276}\\ \midrule
    %

\multicolumn{1}{c|}{\textbf{Technique}}& \multicolumn{1}{c|}{4, 76} & \multicolumn{1}{c|}{10.480} & \multicolumn{1}{c|}{\textless .001} & \multicolumn{1}{c}{.360}\\
    \multicolumn{1}{c|}{\textbf{Distance}} & \multicolumn{1}{c|}{1.552, 29.496}& \multicolumn{1}{c|}{114.720} & \multicolumn{1}{c|}{\textless .001} & \multicolumn{1}{c}{.860}\\ \midrule\midrule
\multicolumn{5}{c}{\textbf{Error Distance}}\\ \midrule
\multicolumn{1}{c|}{\textbf{\noSig{Tech. $\times$ Dist.}}} &\multicolumn{1}{c|}{\noSig{8, 152}}    & \multicolumn{1}{c|}{\noSig{1.446}}    &   \multicolumn{1}{c|}{\noSig{.182}}    &   \multicolumn{1}{c}{\noSig{.071}}\\ \midrule
    \multicolumn{1}{c|}{\textbf{Technique}} & \multicolumn{1}{c|}{4, 76}& \multicolumn{1}{c|}{8.251} & \multicolumn{1}{c|}{\textless .001} & \multicolumn{1}{c}{.303}\\
    \multicolumn{1}{c|}{\textbf{\noSig{Distance}}} & \multicolumn{1}{c|}{\noSig{2, 38}}& \multicolumn{1}{c|}{\noSig{2.826}} & \multicolumn{1}{c|}{\noSig{.072}} & \multicolumn{1}{c}{\noSig{.129}}\\ \midrule 
\end{tabular}
\label{table:Result_With_B}
\vspace{-0.4cm}
\end{table}
\begin{figure}[t]
\centering
\includegraphics[angle=270, width=.45\textwidth]{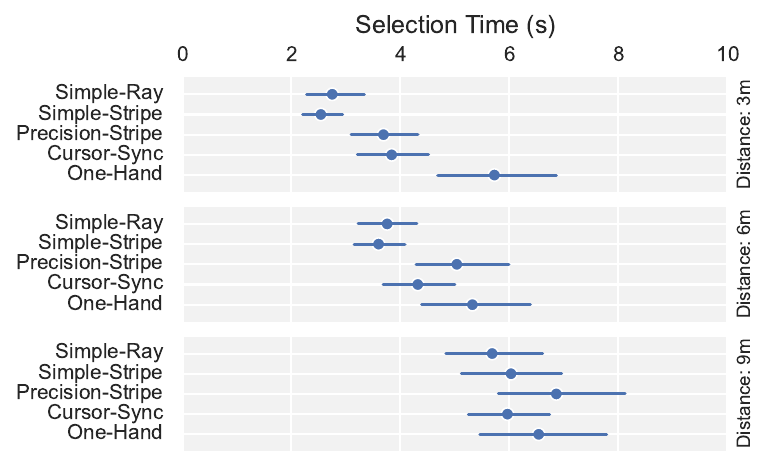}
\caption{
Task 1 - Selection Time results with 95\% CI:
\dhk{
Lower values indicate better and faster performance, whereas higher values signify poorer performance.
}
} 
\label{figure:RT_W/R}
\vspace{-0.4cm}
\end{figure}

\begin{figure}[t]
\centering
\includegraphics[angle=270, width=.45\textwidth]{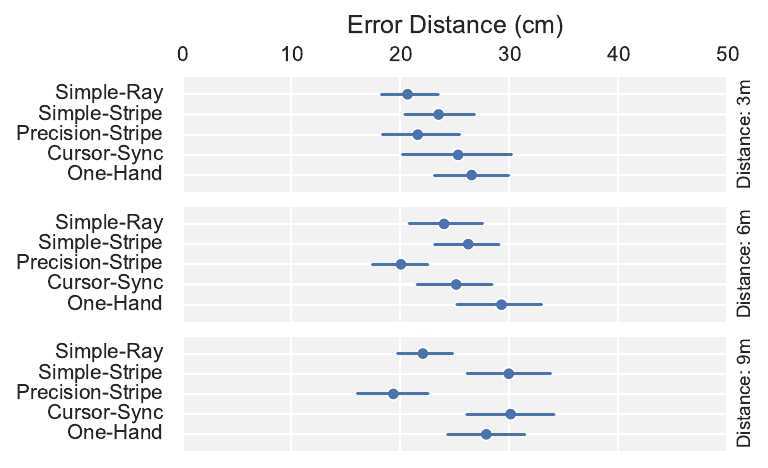}
\caption{
Task 1 - Error Distance results with 95\% CI:
\dhk{Lower values indicate accurate performance, while
higher values are inaccurate. 
}
} 
\label{figure:ED_W/R}
\vspace{-0.3cm}
\end{figure}

This section reports the study results.
We utilize a two-way repeated measures Analysis of Variance (ANOVA) test, with a significance level of 5\%, to analyze selection time, error distance,  NASA-TLX, and UEQ. 
We applied the Geisser-Greenhouse correction to guard against potential violations of the sphericity assumption.
The post-hoc test is conducted by Fisher's least significant differences (LSD) pairwise comparison with $\alpha$ = .05 level for significance.
We employ the median as the representative value for our analysis to mitigate the potential skewness of the collected data.

\subsection{Results: Task 1 \dhk{- \textit{with Reference}}}

\subsubsection{Selection Time}

\begin{table*}[t]
\centering

\caption{Task 1 - NASA-TLX / UEQ Results. The measurements have the statistical differences are only reported. 
Higher values indicate poor performance in terms of effort and frustration but good performance in attractiveness, efficiency, and dependability experience.
}

\begin{tabular} {m{2.2cm}| m{1.7cm}| m{1.0cm}| m{0.8cm}| m{0.7cm} | m{1.4cm} | m{1.4cm}| m{1.4cm} | m{1.4cm}| m{1.4cm}}
\toprule 
  {} & \multicolumn{4}{c|}{\textbf{ANOVA Results}} & \multicolumn{5}{c}{\textbf{Mean (Standard Deviation) by Techniques}}\\ \cmidrule{2-10}
    {}& \multicolumn{1}{c|}{df} & \multicolumn{1}{c|}{F} & \multicolumn{1}{c|}{\textbf{$p$}} & \multicolumn{1}{c|}{\textbf{$\eta$\textsubscript{$p$}\textsuperscript{$2$}}} & {\textbf{Simple-Ray}}  & {\textbf{Simple-Stripe}} & {\textbf{Precision-Stripe}} & {\textbf{Cursor-Sync}} & {\textbf{One-Hand}}\\ \midrule
    \multicolumn{1}{c|}{Effort} & \multicolumn{1}{c|}{4, 76}& \multicolumn{1}{c|}{3.24}& \multicolumn{1}{c|}{.016}& \multicolumn{1}{c|}{.146}& {3.30 (1.72)} & {3.55 (1.79)} & {2.80 (1.67)} & {2.70 (1.81)} & {3.75 (1.71)}\\ 
    \multicolumn{1}{c|}{Frustration Level} & \multicolumn{1}{c|}{2.56,48.75}& \multicolumn{1}{c|}{5.29}& \multicolumn{1}{c|}{.005}& \multicolumn{1}{c|}{.218}& {2.85 (2.03)} & {2.80 (1.82)} & {1.65 (1.14)} & {1.90 (1.48)} & {2.60 (1.93)}\\ \midrule
    \multicolumn{1}{c|}{Attractiveness} & \multicolumn{1}{c|}{2.11, 40.07}& \multicolumn{1}{c|}{3.34}& \multicolumn{1}{c|}{.043}& \multicolumn{1}{c|}{.149}& {0.62 (1.23)} & {0.90 (1.24)} & {1.30 (1.27)} & {1.42 (1.22)} & {0.61 (1.40)}\\ 
    \multicolumn{1}{c|}{Efficiency} & \multicolumn{1}{c|}{4, 76}& \multicolumn{1}{c|}{3.38}& \multicolumn{1}{c|}{.013}& \multicolumn{1}{c|}{.151}& {1.23 (1.09)} & {1.05 (1.21)} & {1.60 (0.84)} & {1.74 (0.87)} & {0.85 (1.27)}\\ 
    \multicolumn{1}{c|}{Dependability} & \multicolumn{1}{c|}{4, 76}& \multicolumn{1}{c|}{4.71}& \multicolumn{1}{c|}{.002}& \multicolumn{1}{c|}{.199}& {0.74 (1.02)} & {0.98 (1.19)} & {1.29 (0.87)} & {1.50 (1.01)} & {0.77 (0.97)}\\ 
    \midrule
\end{tabular}
\label{table:TLX_UEQ_W_B}
\vspace{-0.4cm}
\end{table*}

The selection time results and ANOVA analysis in Task 1 are reported in Figure \ref{figure:RT_W/R} \dhkk{and Table \ref{table:Result_With_B}}.
There is an interaction effect between the technique and distance ($p$ \textless .001).
Further analysis reveals that there are simple effects on technique at 3m (F(1.807, 24.336) = 24.627, $p$ \textless .001, $\eta$\textsubscript{$p$}\textsuperscript{$2$} = .564) and 6m (F(2.626, 49.890) = 10.394, $p$ \textless .001, $\eta$\textsubscript{$p$}\textsuperscript{$2$} = .354). However, there is no simple effect at 9m.
Pairwise comparisons reveal that, at the 3m target position, \dhkk{\textbf{Simple-Ray}} has faster selection time (M = 2.75s, SD = 1.25) than Precision-Stripe (M = 3.69s, SD = 1.45, $p$ \textless .001), Cursor-Sync (M = 3.84s, SD = 1.56, $p$ = .002), and One-Hand (M = 5.73s, SD = 2.6, $p$ \textless .001).
In addition, \dhkk{\textbf{Simple-Stripe}} leads to faster selection time (M = 2.54s, SD = 0.89) than Precision-Stripe ($p$ \textless .001), Cursor-Sync ($p$ = .002), and One-Hand ($p$ \textless .001).
Furthermore, \dhkk{\textbf{One-Hand}} has a significantly faster than Precision-Stripe ($p$ = .002) and Cursor-Sync ($p$ \textless .001).

At 6m, \dhkk{\textbf{Simple-Ray}} has a faster selection time (M = 3.76s, SD = 1.24) than Precision-Stripe (M = 5.04s, SD = 2.04, $p$ = .002) and One-Hand (M = 5.33s, SD = 2.40, $p$ \textless .001).
\dhkk{\textbf{Simple-Stripe}} also has a faster selection time (M = 3.60s, SD = 1.11) than Precision-Stripe ($p$ \textless .001), Cursor-Sync (M = 4.32s, SD = 1.51, $p$ = .027), and One-Hand ($p$ \textless .001).
Moreover, \dhkk{\textbf{Cursor-Sync}} has a faster completion time than the Precision-Stripe ($p$ = .033) and One-Hand ($p$ = .008).

There is a main effects on technique ($p$ \textless .001).
Pairwise comparisons show that \dhkk{\textbf{Simple-Ray}} (M = 4.07s, SD = 2.00) has a faster selection time than Precision-Stripe (M = 5.20s, SD = 2.48, $p$ = .004), Cursor-Sync (M = 4.71s, SD = 1.83, $p$ = .042), and One-Hand (M = 5.87s, SD = 2.58, $p$ \textless .001).
In addition, \dhkk{\textbf{Simple-Stripe}} also results (M = 4.06s, SD = 2.11) faster than Precision-Stripe ($p$ = .009), Cursor-Sync ($p$ = .044), and One-Hand ($p$ \textless .001).
Moreover, \dhkk{\textbf{Cursor-Sync}} is significantly faster than One-Hand ($p$ \textless .001).

A significant main effect is found on distance ($p$ \textless .001).
Pairwise comparisons reveal that \dhkk{\textbf{3m}} (M = 3.71s, SD = 1.98) has a faster selection time than 6m (M = 4.41s, SD = 1.83, $p$ \textless .001) and 9m (M = 6.22s, SD = 2.34, $p$ \textless .001).
In addition, \dhkk{\textbf{6m}} is faster than 9m ($p$ \textless .001).
These results partially support \textbf{H1} and \textbf{H2}.

\subsubsection{Error Distance}

The results are reported in Figure \ref{figure:ED_W/R} and Table \ref{table:Result_With_B}.
A significant main effect is found on technique ($p$ \textless .001).
Pairwise comparisons show that \dhkk{\textbf{Simple-Ray}} has a smaller error distance (M = 22.26cm, SD = 6.77) than Simple-Stripe (M = 26.6cm, SD = 8.21, $p$ \textless .001), Cursor-Sync (M = 26.86cm, SD = 9.86, $p$ = .005), and One-Hand (M = 27.91cm, SD = 8.64, $p$ = .003).
In addition, \dhkk{\textbf{Precision-Stripe}} has a significantly a smaller error distance (M = 20.35cm, SD = 7.32) than Simple-Stripe ($p$ \textless .001), Cursor-Sync ($p$ \textless .001), and One-Hand ($p$ = .002).
These results do not support \textbf{H3} and \textbf{H4}.

\subsubsection{NASA-TLX}

The NASA-TLX results are presented in the upper part of Table \ref{table:TLX_UEQ_W_B}.

\paragraph{Effort}
There is a main effect on Effort ($p$ = .016).
A pairwise comparison shows that \dhkk{\textbf{Precision-Stripe}} has a lower effort rate than Simple-Stripe ($p$ = .014) and One-Hand ($p$ = .016).
In addition, \dhkk{\textbf{Cursor-Sync}} has a lower Effort rate than Simple-Stripe ($p$ = .034) and One-Hand ($p$ = .018).

\paragraph{Frustration}
A main effect on Frustration ($p$ = .005) is found.
Pairwise comparisons reveal that \dhkk{\textbf{Precision-Stripe}} has a lower value than Simple-Ray ($p$ = .006), Simple-Stripe ($p$ = .003), Cursor-Sync ($p$ = .016), and One-Hand ($p$ = .040).
In addition, \dhkk{\textbf{Cursor-Sync}} has a lower Frustration rate than Simple-Ray ($p$ = .040).

There are no significant main effects on Mental Demand, Physical Demand, Performance, and Temporal Demand on technique.
\dhk{
These results are partially support our hypothesis (\textbf{H5}).
}

\subsubsection{UEQ}
The results of the UEQ are presented in the lower part of Table \ref{table:TLX_UEQ_W_B}.
\paragraph{Attractiveness}
There is a main effect on Attractiveness ($p$ = .043).
Pairwise comparisons show that \dhkk{\textbf{Cursor-Sync}} has a higher value than Simple-Ray ($p$ = .021) and Simple-Stripe ($p$ = .047).

\paragraph{Efficiency}
There is a main effect on Efficiency ($p$ = .013).
\dhkk{\textbf{Cursor-Sync}} has higher Efficiency rate than Simple-Stripe ($p$ = .037) and One-Hand ($p$ = .020).
In addition, \dhkk{\textbf{Precision-Stripe}} has a higher Efficiency rate than One-Hand ($p$ = .036).

\paragraph{Dependability}
There is a main effect on Dependability ($p$ = .002). 
Comparisons show that \dhkk{\textbf{Precision-Stripe}} has a higher Dependability rate than Simple-Ray ($p$ = .036) and One-Hand ($p$ = .023).
In addition, \dhkk{\textbf{Cursor-Sync}} has a higher Dependability rate than Simple-Ray ($p$ = .005), Simple-Stirpe ($p$ = .032), and One-Hand ($p$ = .006).

There is no significant difference in Novelty, Perspicuity, and Stimulation among techniques.
These results are partially support our hypothesis (\textbf{H5}).

\subsection{Results: Task 2 \dhk{- \textit{without Reference}}}
\subsubsection{Selection Time}

\begin{table}
\centering
\caption{
Task 2 - ANOVA Analysis Results.
}
\begin{tabular} {m{1.43cm}| m{0.7cm} | m{0.3cm} | m{0.54cm} | m{0.457cm}}
\toprule \multicolumn{5}{c}{\textbf{Selection Time}}\\ \midrule

    {\textbf{}}& \multicolumn{1}{c|}{df} & \multicolumn{1}{c|}{F} & \multicolumn{1}{c|}{\textbf{$p$}} & \multicolumn{1}{c}{\textbf{$\eta$\textsubscript{$p$}\textsuperscript{$2$}}} \\ \midrule

    \multicolumn{1}{c|}{\textbf{Tech $\times$ Dist}} & \multicolumn{1}{c|}{8, 152}& \multicolumn{1}{c|}{13.324}&   \multicolumn{1}{c|}{\textless .001}    &   \multicolumn{1}{c}{.412}\\ \midrule
    %
    
\multicolumn{1}{c|}{\textbf{Technique}} & \multicolumn{1}{c|}{4, 76}& \multicolumn{1}{c|}{88.065} & \multicolumn{1}{c|}{\textless .001} & \multicolumn{1}{c}{.823}\\
    \multicolumn{1}{c|}{\textbf{Distance}} & \multicolumn{1}{c|}{2, 38} & \multicolumn{1}{c|}{56.226} & \multicolumn{1}{c|}{\textless .001} & \multicolumn{1}{c}{.747}\\ \midrule\midrule
\multicolumn{5}{c}{\textbf{Error Distance}}\\ \midrule
\multicolumn{1}{c|}{\textbf{Tech $\times$ Dist}} & \multicolumn{1}{c|}{8, 152}& \multicolumn{1}{c|}{11.634}    &   \multicolumn{1}{c|}{\textless .001}    &   \multicolumn{1}{c}{.380}\\ \midrule
    \multicolumn{1}{c|}{\textbf{Technique}} & \multicolumn{1}{c|}{4, 76}& \multicolumn{1}{c|}{9.294} & \multicolumn{1}{c|}{\textless .001} & \multicolumn{1}{c}{.328}\\
    \multicolumn{1}{c|}{\textbf{Distance}} & \multicolumn{1}{c|}{2, 38}& \multicolumn{1}{c|}{90.853} & \multicolumn{1}{c|}{\textless .001} & \multicolumn{1}{c}{.827}\\ \midrule 

\end{tabular}
\label{table:Result_Without_B}
\vspace{-0.4cm}
\end{table}

\begin{figure}[t]
\centering
\includegraphics[angle=270, width=.45\textwidth]{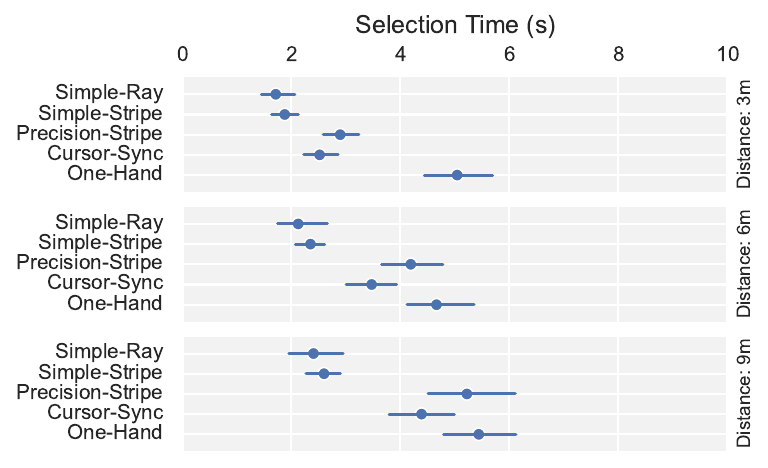}
\caption{
Task 2 - Selection Time results with 95\% CI.
\dhk{
Lower values indicate better/faster performance, whereas higher values signify poorer performance.
}
} 
\label{figure:RT_O/R}
\vspace{-0.3cm}
\end{figure}

The selection time results and ANOVA analysis in Task 2 are reported in Figure \ref{figure:RT_O/R} \dhkk{and Table \ref{table:Result_Without_B}}.
These results show an interaction effect between the technique type and the target position distance ($p$ \textless .001).
Further analysis reveals that there are simple effects on technique at 3m (F(1.944, 36.929) = 72.526, $p$ \textless .001, $\eta$\textsubscript{$p$}\textsuperscript{$2$} = .792), 6m (F(4, 76) = 49.594, $p$ \textless .001, $\eta$\textsubscript{$p$}\textsuperscript{$2$} = .723), and 9m (F(4, 76) = 59.515, $p$ \textless .001, $\eta$\textsubscript{$p$}\textsuperscript{$2$} = .758).
Pairwise comparisons reveal that, at the 3m target position, \dhkk{\textbf{Simple-Ray}} has faster selection time (M = 1.71s, SD = 0.73) than Precision-Stripe (M = 2.90s, SD = 0.77, $p$ \textless .001), Cursor-Sync (M = 2.52s, SD = 0.73, $p$ \textless .001), and One-Hand (M = 5.05s, SD = 1.46, $p$ \textless .001).
In addition, \dhkk{\textbf{Simple-Stripe}} leads to faster selection time (M = 1.88s, SD = 0.60) than Precision-Stripe ($p$ \textless .001), Cursor-Sync ($p$ \textless .001), and One-Hand ($p$ \textless .001).
Furthermore, \dhkk{\textbf{Precision-Stripe}} has a significantly faster selection than One-Hand ($p$ \textless .001).
Lastly, \dhkk{\textbf{Cursor-Sync}} has a faster selection time than Precision-Stripe ($p$ = .037) and One-Hand ($p$ \textless .001).

At 6m, \dhkk{\textbf{Simple-Ray}} has faster selection time (M = 2.13s, SD = 1.06) than Precision-Stripe (M = 4.19s, SD = 1.39, $p$ \textless .001), Cursor-Sync (M = 3.47s, SD = 1.06, $p$ \textless .001), and One-Hand (M = 4.67s, SD = 1.39, $p$ \textless .001).
In addition, \dhkk{\textbf{Simple-Stripe}} leads to faster selection time (M = 2.35s, SD = 0.63) than Precision-Stripe ($p$ \textless .001), Cursor-Sync ($p$ \textless .001), and One-Hand ($p$ \textless .001).
Furthermore, \dhkk{\textbf{Precision-Stripe}} has a significantly faster selection than One-Hand ($p$ = .037).
Lastly, \dhkk{\textbf{Cursor-Sync}} has a faster selection time than Precision-Stripe ($p$ = .014) and One-Hand ($p$ \textless .001).

At 9m, \dhkk{\textbf{Simple-Ray}} has faster selection time (M = 2.41s, SD = 1.22) than Precision-Stripe (M = 5.22s, SD = 1.92, $p$ \textless .001), Cursor-Sync (M = 4.39s, SD = 1.37, $p$ \textless .001), and One-Hand (M = 5.44s, SD = 1.45, $p$ \textless .001).
In addition, \dhkk{\textbf{Simple-Stripe}} leads to faster selection time (M = 2.60s, SD = 0.71) than Precision-Stripe ($p$ \textless .001), Cursor-Sync ($p$ \textless .001), and One-Hand ($p$ \textless .001).
Furthermore, \dhkk{\textbf{Cursor-Sync}} has a faster selection time than Precision-Stripe ($p$ = .011) and One-Hand ($p$ \textless .001).

There are main effects on technique ($p$ \textless .001) and distance ($p$ \textless .001).
Pairwise comparisons show that \dhkk{\textbf{Simple-Ray}} (M = 2.08s, SD = 1.05) has a faster selection time than Precision-Stripe (M = 4.11s, SD = 1.71, $p$ \textless .001), Cursor-Sync (M = 3.46s, SD = 1.31, $p$ \textless .001), and One-Hand (M = 5.05s, SD = 1.45, $p$ \textless .001).
In addition, \dhkk{\textbf{Simple-Stripe}} has a faster result (M = 2.28s, SD = 0.71) than Precision-Stripe ($p$ \textless .001), Cursor-Sync ($p$ \textless .001), and One-Hand ($p$ \textless .001)
Furthermore, \dhkk{\textbf{Simple-Stripe}} has a faster completion time than One-Hand ($p$ \textless .001), and Cursor-Sync has a faster completion time than Precision-Stripe ($p$ = .008) and One-Hand ($p$ \textless .001).

Lastly, \dhkk{\textbf{3m}} (M = 2.81s, SD = 1.50) has faster selection than 6m (M = 3.36s, SD = 1.50, $p$ \textless .001) and 9m (M = 4.01s, SD = 1.87, $p$ \textless .001).
In addition, \dhkk{\textbf{6m}} has faster result than 9m ($p$ \textless .001).
These results support \textbf{H1} and \textbf{H2}.

\subsubsection{Error Distance}

\begin{figure}[t]
\centering
\includegraphics[angle=270, width=.45\textwidth]{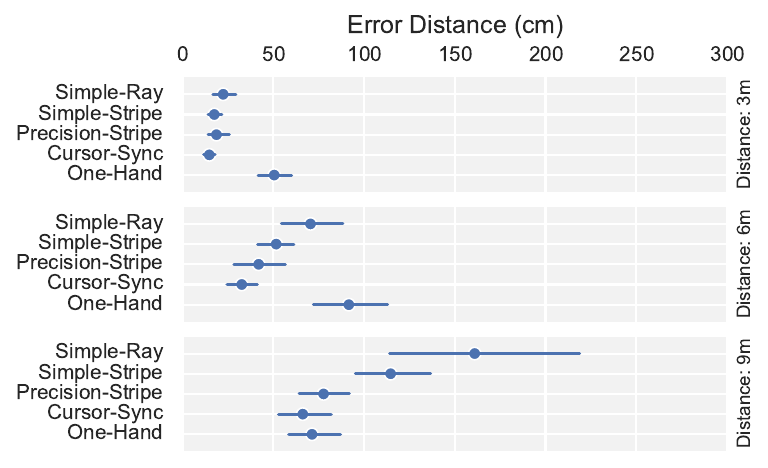}
\caption{
Task 2 - Error Distance results with 95\% CI:
\dhk{
Lower values indicate accurate performance, while
higher values are inaccurate. 
}
} 
\label{figure:ED_O/R}
\vspace{-0.3cm}
\end{figure}

\begin{table*}[t]
\centering
\caption{Task 2 - NASA-TLX / UEQ Results. Only measurements that show statistical differences are reported. 
Higher values indicate poor performance in terms of mental demand and effort but good performance in other measurements.
}
\begin{tabular} {m{2.05cm}| m{1.7cm}| m{1.5cm}| m{0.8cm}| m{0.7cm} | m{1.4cm} | m{1.4cm}| m{1.4cm} | m{1.4cm}| m{1.4cm}}
\toprule 
  {} & \multicolumn{4}{c|}{\textbf{ANOVA Results}} & \multicolumn{5}{c}{\textbf{Mean (Standard Deviation) by Techniques}}\\ \cmidrule{2-10}
    {} & \multicolumn{1}{c|}{df}& \multicolumn{1}{c|}{F} & \multicolumn{1}{c|}{\textbf{$p$}} & \multicolumn{1}{c|}{\textbf{$\eta$\textsubscript{$p$}\textsuperscript{$2$}}} & {\textbf{Simple-Ray}}  & {\textbf{Simple-Stripe}} & {\textbf{Precision-Stripe}} & {\textbf{Cursor-Sync}} & {\textbf{One-Hand}}\\ \midrule
    \multicolumn{1}{c|}{Mental Demand} & \multicolumn{1}{c|}{3.07, 58.34}& \multicolumn{1}{c|}{4.57} & \multicolumn{1}{c|}{.006} & \multicolumn{1}{c|}{.194} & {3.15 (1.66)} & {2.85 (1.79)} & {3.00 (1.69)} & {3.10 (1.97)} & {4.25 (1.94)}\\ 
    \multicolumn{1}{c|}{Performance} & \multicolumn{1}{c|}{4, 76}& \multicolumn{1}{c|}{7.46} & \multicolumn{1}{c|}{\textless .001} & \multicolumn{1}{c|}{.282} & {4.85 (1.76)} & {4.95 (1.61)} & {5.35 (1.57)} & {4.80 (1.61)} & {3.65 (1.50)}\\
    \multicolumn{1}{c|}{Effort} & \multicolumn{1}{c|}{4, 76}& \multicolumn{1}{c|}{5.15} & \multicolumn{1}{c|}{\textless .001} & \multicolumn{1}{c|}{.213} & {3.25 (1.89)} & {2.80 (1.40)} & {3.25 (1.80)} & {2.75 (1.68)} & {4.00 (1.89)}\\ \midrule
    \multicolumn{1}{c|}{Attractiveness} & \multicolumn{1}{c|}{2.39, 45.32}& \multicolumn{1}{c|}{8.37} & \multicolumn{1}{c|}{\textless .001} & \multicolumn{1}{c|}{.306} & {1.18 (1.11)} & {1.14 (1.51)} & {1.58 (1.12)} & {1.31 (1.16)} & {0.18 (1.57)}\\ 
    \multicolumn{1}{c|}{Efficiency} & \multicolumn{1}{c|}{2.30, 43.79}& \multicolumn{1}{c|}{5.42} & \multicolumn{1}{c|}{.006} & \multicolumn{1}{c|}{.222} & {1.59 (0.80)} & {1.58 (1.11)} & {1.75 (0.86)} & {1.81 (0.97)} & {0.79 (1.24)}\\ 
    \multicolumn{1}{c|}{Perspicuity} & \multicolumn{1}{c|}{2.49, 47.23}& \multicolumn{1}{c|}{6.42} & \multicolumn{1}{c|}{.002} & \multicolumn{1}{c|}{.253} & {2.31 (0.81)} & {2.04 (1.28)} & {2.11 (1.02)} & {2.00 (1.23)} & {1.10 (1.32)}\\
    \multicolumn{1}{c|}{Dependability} & \multicolumn{1}{c|}{2.22, 42.17}& \multicolumn{1}{c|}{6.89} & \multicolumn{1}{c|}{.002} & \multicolumn{1}{c|}{.266} & {1.09 (0.94)} & {1.24 (0.88)} & {1.39 (0.84)} & {1.33 (0.97)} & {0.37 (1.21)}\\
    \multicolumn{1}{c|}{Stimulation} & \multicolumn{1}{c|}{4, 76}& \multicolumn{1}{c|}{4.26} & \multicolumn{1}{c|}{.004} & \multicolumn{1}{c|}{.183} & {0.86 (1.23)} & {0.88 (1.48)} & {1.21 (1.19)} & {1.20 (0.97)} & {0.23 (1.34)}\\ \midrule
\end{tabular}
\label{table:TLX_UEQ_Without_B}
\vspace{-0.4cm}
\end{table*}

The error distance results with each condition are reported in Figure \ref{figure:ED_O/R}.
There is an interaction effect between technique and target position distance ($p$ \textless .001).
Further examination shows significant simple effects of the techniques at distances of 3m (F(2.505, 47.596) = 24.371, $p$ \textless .001, $\eta$\textsubscript{$p$}\textsuperscript{$2$} = .562), 6m (F(4, 76) = 12.079, $p$ \textless .001, $\eta$\textsubscript{$p$}\textsuperscript{$2$} = .389), and 9m (F(1.755, 33.350) = 9.300, $p$ \textless .001, $\eta$\textsubscript{$p$}\textsuperscript{$2$} = .329).
Pairwise comparisons show that, at the 3m target position, \dhkk{
\textbf{Simple-Ray} (M = 22.50cm, SD = 14.30), \textbf{Simple-Stripe} (M = 17.61cm, SD = 9.25), \textbf{Precision-Stripe} (M = 18.80cm, SD = 14.69) and \textbf{Cursor-Sync} (M = 14.83cm, SD = 7.37) have a significantly smaller error distance than One-Hand (M = 50.60cm, SD = 21.37, $p$ \textless .001)
In addition, \textbf{Cursor-Sync} ($p$ = .041) has a smaller error distance than the Simple-Ray technique.
}

At 6m, \dhkk{\textbf{Simple-Stripe}} has \dhkk{smaller error} results (M = 51.68cm, SD = 23.42) than Simple-Ray (M = 70.58cm, SD = 40.13, $p$ = .025) and One-Hand (M = 91.70cm, SD = 49.75, $p$ = .001).
Moreover, \dhkk{\textbf{Cursor-Sync} has a smaller error result} (M = 32.72cm, SD = 19.74) than Simple-Ray ($p$ \textless .001), Simple-Stripe ($p$ = .014), and One-Hand ($p$ \textless .001) techniques.

\dhkk{At 9m, \textbf{Precision-Stripe} (M = 77.78cm, SD = 32.08), \textbf{Cursor-Sync} (M = 66.33cm, SD = 35.23), and \textbf{One-Hand} (M = 71.42cm, SD = 32.26) techniques has smaller error results than the Simple-Ray (Precision-Stripe: $p$ = .002; Cursor-Sync: $p$ \textless .001; One-Hand: $p$ = .006) and Simple-Stripe (M = 114.71cm, SD = 45.74, Precision-Stripe: $p$ = .013; Cursor-Sync: $p$ = .003; One-Hand: $p$ = .001) techniques.
}

There are main effects on technique ($p$ \textless .001) and distance ($p$ \textless .001).
Pairwise comparisons show that \dhkk{\textbf{Precision-Stripe}} has a smaller error distance (M = 46.21cm, SD = 37.21) than Simple-Ray (M = 84.70cm, SD = 91.63, $p$ \textless .001) and One-Hand (M = 71.24cm, SD = 39.57, $p$ = .002).
In addition, \dhkk{\textbf{Cursor-Sync}} has a smaller error result (M = 37.96cm, SD = 31.72) than Simple-Ray ($p$ \textless .001), Simple-Stripe (M = 61.34cm, SD = 50.23, $p$ = .004), Precision-Stripe ($p$ = .046), and One-Hand ($p$ \textless .001).
Lastly, the participants complete the task with a smaller error when a target position is \dhkk{\textbf{3m}} (M = 24.87cm, SD = 19.20) than 6m (M = 57.74cm, SD = 40.54, $p$ \textless .001) and 9m (M = 98.25cm, SD = 70.57, $p$ \textless .001), and also, the result from \dhkk{\textbf{6m}} is smaller than from 9m ($p$ \textless .001).
These results partially support \textbf{H3} and \textbf{H4}.

\subsubsection{NASA-TLX}

The NASA-TLX results are shown in the upper part of the Table \ref{table:TLX_UEQ_Without_B}.

\paragraph{Mental Demand}
There is a main effect on Mental Demand ($p$ = .006).
Pairwise comparisons show that \dhkk{\textbf{One-Hand}} has a higher Mental Demand rate than Simple-Ray ($p$ = .005), Simple-Stripe ($p$ = .001), Precision-Stripe ($p$ = .008), and Cursor-Sync ($p$ = .012).

\paragraph{Performance}
 In addition, there is a significant main effect on Performance ($p$ \textless .001).
 \dhkk{\textbf{One-Hand}} has lower performance rate than Simple-Ray ($p$ \textless .001), Simple-Stripe ($p$ \textless .001), Precision-Stripe ($p$ \textless .001), and Cursor-Sync ($p$ \textless .001).

\paragraph{Effort}
There is a main effect on Effort ($p$ \textless .001).
\dhkk{\textbf{One-Hand}} has higher Effort rate than Simple-Ray ($p$ = .021), Simple-Stripe ($p$ = .006), and Cursor-Sync ($p$ = .002).
However, there is no difference between One-Hand and Precision-Stripe, and among Simple-Ray, Simple-Stripe, and Cursor-Sync.

There is no significant main effect on Frustration, Physical Demand, and Temporal Demand.
\dhk{
These results are partially support our hypothesis (\textbf{H5}).
}
\subsubsection{UEQ}

The UEQ results are presented in the bottom part of the Table \ref{table:TLX_UEQ_Without_B}.

\paragraph{Attractiveness}
There is a main effect on Attractiveness ($p$ \textless .001).
Pairwise comparisons show that \dhkk{\textbf{One-Hand}} has a lower Attractiveness rate than Simple-Ray ($p$ \textless .001), Simple-Stripe ($p$ = .012), Precision-Stripe ($p$ \textless .001), and Cursor-Sync ($p$ = .003).
In addition, \dhkk{\textbf{Simple-Stripe}} has a lower Attractiveness rate than Precision-Stripe ($p$ = .022).

\paragraph{Efficiency}
There is a main effect on Efficiency ($p$ = .006).
\dhkk{\textbf{One-Hand}} has a lower Efficiency rate than Simple-Ray ($p$ = .011), Simple-Stripe ($p$ = .039), Precision-Stripe ($p$ = .005), and Cursor-Sync ($p$ = .008).

\paragraph{Perspicuity}
There is a main effect on Perspicuity ($p$ = .002).
\dhkk{\textbf{One-Hand}} has a lower perspicuity rate than Simple-Ray ($p$ \textless .001),  Simple-Stripe ($p$ = .007), Precision-Stripe ($p$ = .002), and Cursor-Sync ($p$ = .026).

\paragraph{Dependability}
Moreover, there is a main effect on Dependability ($p$ = .002).
\dhkk{\textbf{One-Hand}} has a lower Dependability rate than Simple-Ray ($p$ = .029), Simple-Stripe ($p$ = .005), Precision-Stripe ($p$ = .002), and Cursor-Sync ($p$ = .006). 

\paragraph{Stimulation}
Lastly, there is a main effect on Stimulation ($p$ = .004).
\dhkk{\textbf{One-Hand}} has a lower Stimulation rate than Simple-Ray ($p$ = .020), Precision-Stripe ($p$ = .011) and Cursor-Sync ($p$ = .006).
However, there is no significant difference between One-Hand and Simple-Stripe.

The Novelty of each technique is not significantly different in this task.
\dhk{
These results are partially support our hypothesis (\textbf{H5}).
}

%% file: ANOVA_Result_contents/5_discussion.tex
\section{Discussion}
In this section, we discuss our findings from the study on mid-air selection, its limitations, and our future research directions.

\paragraph{Simple-Ray outperformed others in the \textit{with Reference} task}
When a reference object was given for mid-air selection \dhk{(Task 1)},  Simple-Ray outperformed Simple-Stripe, Precision-Stripe, Cursor-Sync, and One-Hand techniques in terms of task completion time and error distance, \dhk{supporting our hypothesis (\textbf{H1})}.
Specifically, when the target was located 3m away, Simple-Ray and Simple-Stripe achieved faster completion times compared to the others.
However, with an increase in target distance, the performance differences between these techniques became negligible.
Furthermore, Simple-Ray demonstrated smaller error distances than other techniques, except for Precision-Stripe, \dhk{rejecting \textbf{H4}}.

An additional noteworthy observation is that, on the whole, One-Hand underperformed compared to the bimanual methods.
However, as the target distance increases, its performance aligns more closely with that of the bimanual techniques.
\dhk{
\paragraph{
The selection support feature could improve accuracy in the \textit{without Reference} task.
}
In Task 2, Precision-Stripe and Cursor-Sync were faster than One-Hand at 3m and 6m yet exhibited no significant performance differences at 9m.
However, those two techniques are slower than Simple-Ray and Simple-Stripe.
This finding indicates that the support features for precise selection can not significantly decrease the time spent for selecting (denying \textbf{H2}).
However, Precision-Stripe and Cursor-Sync exhibited smaller error distances than their counterparts, which aligns with our hypothesis that support features can improve accuracy (supporting \textbf{H3}).
}
%
%
\dhk{ 
\paragraph{
Selection speed and accuracy are traded off depending on a task type.
}
The Task 2 results show a trade-off between selection speed and accuracy.
Precision-Stripe and Cursor-Sync exhibited better accuracy than Simple-Ray and Simple-Stripe while Simple-Ray and Simple-Stripe showed faster selection speeds than Precision-Stripe and Cursor-Sync.
In contrast, from the Task 1 results, Simple-Ray demonstrated faster selection speed and more accuracy than other techniques.
From this dissimilarity, we can guess that when a selectable point is limited (Task 1), supporting features, such as a selected stripe on Precision-Stripe or a synced cursor on Sync-Cursor, can even decrease the selection performance.
}
\paragraph{
Precision-Stripe and Cursor-Sync significantly improve the user experience in the presence of a reference object.}
The NASA-TLX and UEQ results show that both Precision-Stripe and Cursor-Sync provide less workload and better user experiences \dhk{than the other bimanual techniques and One-Hand}.
Specifically, participants reported that both techniques demand lower effort and induce less frustration in tasks involving a reference object.
The result is partially foreseen, the support features can provide less workload, as we hypothesized (supporting \textbf{H5}).
However, the visual aid, which is in Simple-Stripe, can not lead to a significant effect on the workload.
Furthermore, both Precision-Stripe and Cursor-Sync were rated higher in terms of attractiveness, efficiency, and dependability.
This result also indicates the aligned result with the support feature effect for the workload.
From these results, we can guess that only the visual aid can not improve the user experiment, and it needs to support features that facilitate the precision selection.
\dhk{
\paragraph{
Visual aids between users and positions can improve their experiences.
}
In Task 2, the NASA-TLX and UEQ results indicate that the bimanual techniques require less effort and mental demand and surpass the subjective performance of One-Hand.
However, we found no clear difference between the bimanual techniques.
This finding suggests that a ray from the dominant hand side during the target red sphere was shown when using the bimanual technique, which improves the user experience (supporting \textbf{H5}).
Interestingly, that user experience is independent of the type of ray and the actual performance of techniques.
}

\subsection{Limitations}

\dhk{
This section discussed the limitations of the techniques and study. First of all, despite incorporating visual aids and supporting features into the bimanual technique, the error distance remains in the range of tens of centimeters. While this inaccuracy may be acceptable in specific contexts, such as pointing to a distant mid-air position, it poses challenges for general applications, particularly in selecting a destination for navigation.
Consequently, achieving accurate and reliable mid-air selection for navigation remains an open question for future research.}


\dhk{
Secondly, the study only focuses solely on a specific position, 50cm in front of the reference object. 
However, in practice scenarios, users may need to select positions in various directions related to the reference object which could yield different results.
Additionally, in Task 2, it is challenging to determine whether the observed errors are from participants' limited depth perception or the inherent nature of the selection technique.
}

\dhk{
Finally, our study results might be biased due to the specific age group (18 to 22) of the participants, who mostly had prior VR experience. Different demographic groups may yield varying results.}

%% file: ANOVA_Result_contents/6_conclusion.tex
\section{Conclusion}

In this paper, we \hdy{identified and evaluated} four bimanual and one unimanual selection techniques designed \hdy{for mid-air selection}: 
Simple-Ray, Simple-Stripe, Precision-Stripe, and Cursor-Sync.
We evaluated their usability and efficiency in comparison to the One-Hand technique, focusing on two distinct types of tasks: selecting a mid-air position with a reference object and selecting a position without such a reference.
This study delineated the unique advantages of several techniques tailored to two specific task categories, highlighting notable enhancements in task performance and user experience relative to the one-hand mid-air selection technique, particularly for targets at moderate distances.
We present these findings while recognizing the study's limitations and propose directions for future research to explore these constraints further.